\documentclass[11pt]{article}
\title{New Results on the DMC Capacity and Renyi's Divergence}

\author{Yi~Janet Lu\thanks{The author is currently paying an invited visit at EPFL, Switzerland.}\\
Department of Informatics, University of Bergen, 5020 Norway\\
\url{Yi.JANET.Lu@gmail.com}
}

\usepackage{amsmath}
\usepackage{amssymb}
\usepackage{url,algorithmic}
\usepackage{epsfig,graphics,color}
\usepackage{algorithmic,algorithm}

\usepackage[caption=false,font=footnotesize]{subfig}

\newtheorem{conjecture}{Conjecture}
\newtheorem{remark}{Remark}
\begin{document}
\maketitle

\begin{abstract}
This work is part of a project ``Walsh Spectrum Analysis and the Cryptographic Applications''. 
The project initiates the study of finding the largest (and/or significantly large) Walsh coefficients as well as the index positions of an unknown distribution by random sampling. 
This proposed problem has great significance in cryptography and communications.

In early 2015, Yi JANET Lu first constructed novel imaginary channel
transition matrices and introduced Shannon's channel coding problem
to statistical cryptanalysis.
For the first time,
the channel capacity results of well-chosen transition matrices,
which might be impossible to calculate traditionally,
become of greatest research focus.
For a few Discrete Memoryless Channels (DMCs), it is known that the capacity can be computed analytically; 
in general, there is no closed-form solution. 
This work is concerned with analytical results of channel capacity in the new setting. 
We study both the Blahut-Arimoto algorithm (which gave the first numerical solution historically) and the most recent results [Sutter et al'2014] for the transition matrix of $N\times M$. 
For an $\epsilon$-approximation (i.e., the desired absolute accuracy of the approximate solution) of the capacity, the former has the computational complexity $ O(MN^2 \log N/\epsilon) $, while the latter has the complexity 
$ O(M^2N\sqrt{\log N}/\epsilon) $. 
We also study the relation of Renyi's divergence of degree $1/2$ and the generalized channel capacity 
of degree $1/2$.
~\\
\noindent
\textbf{Keywords:}
DMC,
Channel capacity,
Blahut-Arimoto algorithm,
Transition matrix construction,
Statistical cryptanalysis,
Renyi's divergence. 
\end{abstract}

\section{Introduction}
Inspired and influenced by the greatest innovative idea of compressive sensing (cf. \cite{cs_lecture_notes2007}), 
Yi JANET Lu in early 2015 first constructed imaginary channel transition matrices and 
introduced Shannon's channel coding problem (cf. \cite{it-book}) to statistical cryptanalysis 
(cf. \cite{vaudenay_statistical_cryptanalysis}). 
This result \emph{surprisingly} gives a \emph{perfect} answer to the \emph{key} question in cryptanalysis 
\emph{for the first time},
that is,
what is the minimum number of data samples to distinguish one biased distribution from the uniform
distribution?
In this work, we will study the DMC capacity in this new setting. 
In particular,
we study and implement the famous Blahut-Arimoto Algorithm 
in order to calculate the reference value of the DMC capacity in our new setting.
Then,
we did analysis on the novel non-symmetric binary channel,
which plays a crucial role exclusively in statistical cryptanalysis.
We gave the closed-form capacity estimate,
and compare with the results of Blahut-Arimoto Algorithm.
We show that our closed-form capacity estimate is very close
and the well-known crypto estimate formula needs to be updated accordingly.
Further, we are the first to discover another estimated formula by 
Renyi's divergence of degree 1/2 is very precise.
Our work is extended to channels of two input symbols and $M$ output symbols.

\section{Preliminaries on the Blahut-Arimoto Algorithm}
Due to independent works of [Arimoto'1972] and [Blahut'1972],
the famous Blahut-Arimoto algorithm is known to efficiently calculate the
capacity for the discrete memoryless channel (DMCs).
Below, we present the Blahut-Arimoto algorithm\footnote{We called it BA algorithm in short.} in pseudo-codes (see
Fig.~\ref{Fig_alg1}),
which calculates the capacity of arbitrary transition matrices of size $2\times M$
(note that by convention, the notation $Q_{k|j}$ is used to denote the
probability of receiving the $k$-th output symbol when the $j$-th input symbol was transmitted).
This is the \emph{best} algorithm so far to calculate the DMC capacity for
 transition matrix sizes $N\times M$ with $N<M$.
For the desired absolute accuracy $\epsilon$ of the approximate solution,
the algorithm has the computational complexity
$ O\Bigl(MN^2 (\log N)/\epsilon\Bigr) $,
that is,
$ O\Bigl(4M(\log 2)/\epsilon\Bigr) $.

\begin{figure}[h]
\begin{algorithmic}[1]
\REQUIRE ~\\
$Q_{k|j}$: transition matrix of size $2\times M$ \\
 $(p_0,p_1)$: input distribution vector \\
 $\epsilon:$ the desired absolute accuracy
\STATE initialize the values of $Q_{k|j}$ and $p_0,p_1$
\REPEAT
\STATE $c_0 \leftarrow \exp\bigl(\sum_{k=0}^{M-1}Q_{k|0}\log \frac{Q_{k|0}}{p_0Q_{k|0} + p_1Q_{k|1}}\bigr)$
\STATE $c_1 \leftarrow \exp\bigl(\sum_{k=0}^{M-1}Q_{k|1}\log \frac{Q_{k|1}}{p_0Q_{k|0} + p_1Q_{k|1}}\bigr)$
\STATE $I_L \leftarrow \log(p_0c_0 + p_1c_1)$
\STATE $I_U \leftarrow \log \, \max(c_0, c_1)$
\STATE update $p_0$ by ${p_0c_0}/{\bigl(p_0c_0 + p_1c_1\bigr)}$
\STATE update $p_1$ by ${p_1c_1}/{\bigl(p_0c_0 + p_1c_1\bigr)}$
\UNTIL{ $|I_U - I_L| < \epsilon$ }
\STATE output $I_L$
\end{algorithmic}
\caption{DMC Capacity Calculation Pseudo-codes of Blahut-Arimoto Algorithm}\label{Fig_alg1}
\end{figure}

We point out that when 
1) 
$M$ is not a power of two,
2)
the transition matrix contains \emph{strict zero points},
3) $M$ is very big (e.g., $2^{64}$),
 implementation of BA algorithm is a \emph{delicate} issue.
Particularly,
the well-known variable type double 
does not fit to represent $Q_{k|j}$, $p_0, p_1$.
Nonetheless, we begin with binary channels in next section.

\section{Binary Channels}
First, we recall the well-known capacity result for binary symmetric channels (BSC) with crossover probability $p$,
that is, the transition matrix is of the form
\[
\left(
\begin{array}{cc}
1-p & p \\
p & 1-p \\
\end{array}
\right)
\,
.
\]
Let $p=(1-d)/2$, the capacity $C$ (see \cite{it-book}) is strictly equal to 
\[
C = 1 - H(p) \text{ (binary bits/transmission)},
\]
where the binary entropy function $H(p) \stackrel{\text{def}}{=} - p\log_2 (p) - (1-p)\log_2(1-p)$.
For convenience, in units of natural logarithm bits\footnote{From now on, the capacity is always in units of natural logarithm bits rather than binary bits unless mentioned explicitly otherwise.}, we can rewrite
\begin{equation}
\label{Formula_theory_bsc}
C = 
\log(2) + p\log(p) + (1-p)\log(1-p) .
\end{equation}
Meanwhile, we note that
the walsh-hadamard transform of each row of the transition matrix is a matrix of the form
\[
\left(
\begin{array}{cc}
\textcolor{blue}{1} & +d \\
\textcolor{blue}{1} & -d \\
\end{array}
\right)
\,
.
\]
Note that the leading one (in blue) of each row is a \emph{trivial} nonzero coefficient.

\subsection{The Non-Symmetric Binary Channel}
In early 2015, Yi~JANET Lu for the first time constructed a non-symmetric binary channel,
which has the transition matrix of the following form
\[
\left(
\begin{array}{cc}
1-p & p \\
1/2 & 1/2 \\
\end{array}
\right) \, .
\]
It proves the most interesting in cryptography. Again,
the walsh-hadamard transform of each row of the matrix is of the form
\[
\left(
\begin{array}{cc}
\textcolor{blue}{1} & d \\
\textcolor{blue}{1} & 0 \\
\end{array}
\right)
\,
.
\]
This addresses one basic problem in cryptography,
which aims at using the minimum number of samples to distinguish the sequence of i.i.d. \emph{biased} binary bits from a
truly random sequence of equal length.
Let $p = (1-d)/2$. 
Yi~Janet Lu demonstrated that when $d$ is small, 
the capacity $C$ of this channel can be approximated\footnote{see Appendix for the proof.} by
\begin{equation}
\label{Formula_Est_nonsymmetric_binary_channel}
C \approx d^2 /\bigl(8\log 2\bigr) .
\end{equation}
\begin{remark}
This binary channel construction can be used to answer the
(non-)existence of the walsh-hadamard coefficient of a distribution over the binary space, 
which is no smaller than a threshold value (i.e., $|d|$) in absolute value 
using the fixed number of random samples.
\end{remark}

\subsection{Estimate by Renyi's Information Divergence}
We explore the quantitative relation between the capacity and the Renyi's information divergence as stated
by the conjecture below:
\begin{conjecture}
Let $Q,U$ be a non-uniform distribution and a uniform distribution over the support of cardinality $2^n$.
Let the matrix of $T$ consist of two rows $Q,U$ and $2^n$ columns.
We have the following relation between Renyi's divergence of degree $1/2$ and the generalized channel capacity of degree $1/2$
 (i.e., the standard Shannon's channel capacity),
\[
D_{1/2}(Q\|U) = 2\cdot C_{1/2}(T).
\]
\end{conjecture}

Recall that Renyi's information divergence (see \cite{renyi}) of order $\alpha = 1/2$ 
of distribution $P$ from another distribution $Q$ on a finite set $\mathcal{X}$ is defined as
\begin{equation}
D_{\alpha} (P \| Q) \stackrel{\text{def}}{=} 
  \frac{1}{\alpha - 1} \log \sum_{x \in \mathcal{X}} P^{\alpha}\bigl( x \bigr) Q^{1-\alpha}\bigl( x \bigr) .
\end{equation}
So, with $\alpha = 1/2$,
\begin{equation}
D_{1/2} (P \| Q) \stackrel{\text{def}}{=}  (-2) \log \sum_{x \in \mathcal{X}} \sqrt{P(x)Q(x)} .
\end{equation}

According to~\cite{renyi}, taking limit as $\alpha \rightarrow 1$,
 Kullback-Leibler information divergence is recovered as information divergence of order
$\alpha = 1$.

\subsection{Numerical Results}
We list the calculated numerical values of BA algorithm outputs in Appendix, Table~\ref{T1} (for the non-symmetric binary channel) 
and Table~\ref{T0} (for BSC) with $\epsilon=0.0001$.
In both tables, we compare the capacity results with the classical crypto estimates 
as well as using Renyi's quantity.
Fig. (1.a), Fig. (1.b) show the results corresponding to Table~\ref{T1} and Table~\ref{T0} respectively.
Our new founding is that
when $d$ is small, using Renyi's quantity is \emph{surprisingly accurate} AND
the classical crypto estimate is slightly higher.

\begin{figure}[!t]
\centering
\subfloat[][]{\includegraphics[scale=.4]{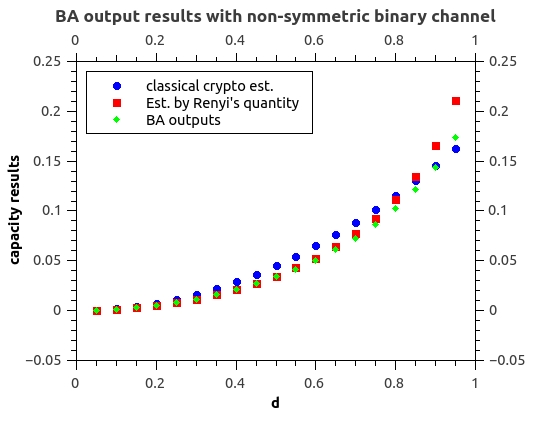}}
\qquad
\subfloat[][]{\includegraphics[scale=.4]{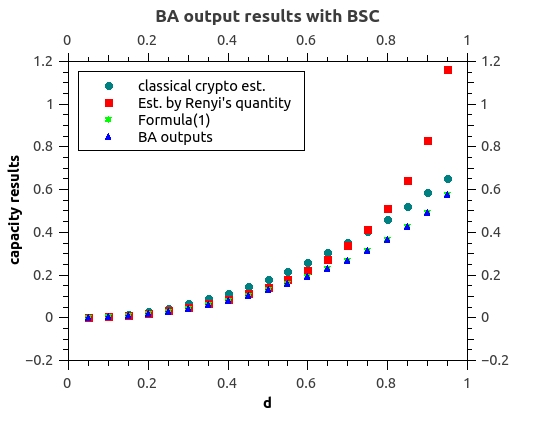}}
\caption{Capacity Results for Non-symmetric Binary Channels (Left) and BSC (Right)}
\label{Fig_binary_channels}
\end{figure}

\begin{remark}
In cryptanalysis, the bias value (i.e., $d$ herein) of the biased bit (together with the corresponding mask $m$)
is \emph{obtained} by \emph{manual analysis} before hand,
which is usually a very hard task.
For a distribution $D$ over the binary vector space of dimension $n$,
which can be defined over potentially larger number of input states,
it is critical for linear cryptanalysis to
find some large walsh-hadamard coefficient $d$ of $D$ together with the mask $m$, 
i.e., $\widehat{D}(m) = d$ is large in absolute value.
\end{remark}

\section{Our Results with $M = 256$}
We decide to run our experiments to compute the capacity with $M = 256, \epsilon=0.0001$.
We choose the transition matrix $Q$ such that $Q_{k|j=1}$ is 
a uniform distribution over the binary vector space of dimension $8$,
and $Q_{k|j=0}$ has the sparsity $k=1$ nonzero walsh-hadamard coefficients $d$ except at the zero point.
We plot the capacity results by BA algorithm in Fig. (2.a) against the 
classical crypto estimate using $k\cdot d^2/(8\log 2)$.
Similarly,
in Fig. (2.b) and Fig. (2.c)
we plot for the sparsity $k=2, 4$ respectively.
Note that $Q_{k|j=0}$ has the sparsity $k$ nonzero walsh-hadamard coefficients with same absolute value $d$.
Fig. (2.d) compares the sparsity $k=1, 2, 4$.
Generally speaking,
the classical crypto esitmate is quite close to the theoretical value.
In the full version of the paper (see \cite{me_capacity_submission}),
we will present more results and discussions when $M$ is not a power of two, $M$ is larger than $2^{30}$
and $k$ takes more choices of values.

\begin{figure}[!t]
\centering
\subfloat[][]{\includegraphics[scale=.4]{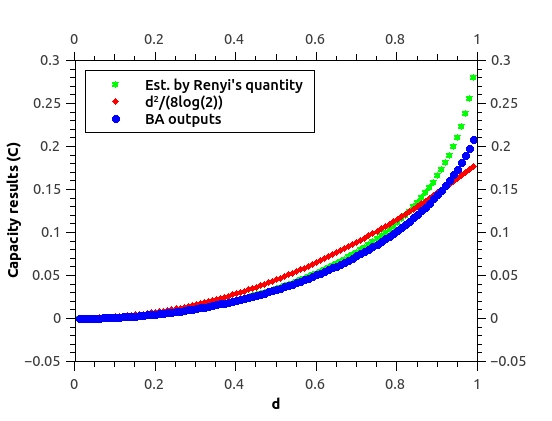}}
\qquad
\subfloat[][]{\includegraphics[scale=.4]{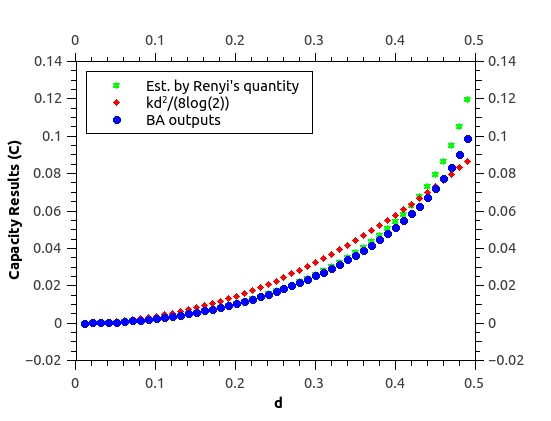}} \\
\medskip
\subfloat[][]{\includegraphics[scale=.4]{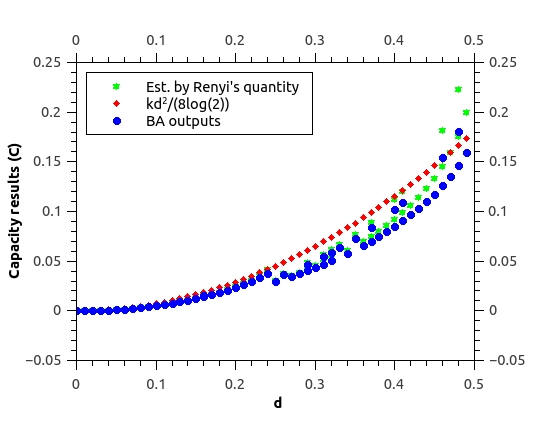}}
\qquad
\subfloat[][]{\includegraphics[scale=.4]{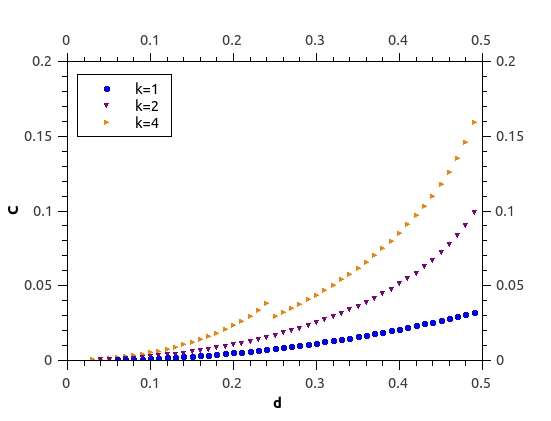}}
\caption{Capacity Results for $M=256,\epsilon = 0.0001$ and $k=1$ (Top Left), $k=2$(Top Right), $k=4$(Bottom Left) and Comparison of BA outputs (Bottom Right).}
\label{Fig_M_256}
\end{figure}

\subsection{Statistical Distinguisher to Solve Shannon's Channel Coding Problem for Our Transition Matrices}
Following the work of statistical cryptanalysis (see~\cite{vaudenay_statistical_cryptanalysis}),
we present the algorithm of the statistical distinguisher in Fig.~\ref{Fig_alg2}.
This can be seen as the answering machine
that solves the Shannon's channel coding problem 
with the transition matrix $T$ of $2\times M$. And
the matrix consists of two rows corresponding to 
the biased distribution and the uniform distribution\footnote{So $D_0(b) = 1/M$ for all $b$.} respectively.

\begin{figure}[h]
\begin{algorithmic}[1]
\REQUIRE ~\\
$n = 8$, $M = 2^n$ \\
$D_0$: the uniform distribution over $n$ bits \\
$D_{A}$: the biased probability distribution of the $n$-bit vector $A$ \\
$N$: sample number \\
$B_0, B_1, \ldots, B_N$: blocks of i.i.d. $n$-bit samples all from $D_A$ or $D_0$
\STATE initialize counters $u[0], u[1],\ldots,u[M-1]$ to zeros 
\FOR{$t = 0, 1, \ldots, N$}
\STATE increment $u[B_t]$
\ENDFOR
\IF{ $\sum_b u[b]\cdot\log\Bigl( D_A(b) / D_0(b)\Bigr) > 0$}
\STATE accept $D_A$ as the source
\ELSE
   \STATE accept $D_0$ as the source
\ENDIF
\end{algorithmic} 
\caption{The Classical Statistical Distinguisher }
\label{Fig_alg2}
\end{figure}

\section*{Appendix A: Proof of Capacity Estimate for the Non-Symmetric Binary Channel}

We now propose a simple method to give a closed-form estimate $C$ (when $d$ is small) for our binary channel.
As $I(X;Y)= H(Y)-H(Y|X)$, we first compute $H(Y)$ by
\begin{equation}\label{E_tmp1}
H(Y) = H\Bigl( p_0(1-p_e)+ (1-p_0)\times \frac{1}{2} \Bigr),
\end{equation}
where $p_0$ denotes $p(x=0)$ for short.
Next, we compute % 
\begin{equation}
H(Y|X) = \sum_x p(x)H(Y|X=x) 
= p_0\Bigl( H(p_e)-1 \Bigr) +1.\label{E_tmp2}
\end{equation}
Combining (\ref{E_tmp1}) and (\ref{E_tmp2}), we have
\[
I(X;Y) = H\Bigl( p_0\times \frac{1}{2} - p_0p_e + \frac{1}{2} \Bigr) - 
  p_0H(p_e) + p_0 -1.
\]
As $p_e=(1-d)/2$, we have
\[
I(X;Y)
=H(\frac{1 + p_0d}{2}) -p_0\Bigl( H(\frac{1-d}{2})-1 \Bigr) - 1.
\] 
For small $d$,
we use the following result to continue
\begin{equation}\label{approx_H_special}
H\Bigl(\frac{1+d}{2}\Bigr) = 1 - d^2 /(2\log 2) + O(d^4) .
\end{equation}
\begin{equation}\label{E_tmp3}
I(X;Y) = -\, \frac{p_0^2 d^2}{2\log 2} - p_0\Bigl( H(\frac{1-d}{2})-1 \Bigr) + O(p_0^4 d^4).
\end{equation}
 Note that
 the last term $O(p_0^4 d^4)$ on the right side of (\ref{E_tmp3}) is ignorable.
 Thus, $I(X;Y)$ is estimated to approach the maximum when
\[
p_0 = -\,\frac{ H(\frac{1-d}{2})-1 }{d^2/(\log 2)} \approx \frac{d^2/(2\log 2)}{d^2/(\log 2)}=\frac{1}{2}.
\]
Consequently, we estimate the channel capacity (\ref{E_tmp3}) by
\begin{equation*}
C \approx -\,\frac{1}{4}d^2/(2\log 2) +\frac{1}{2} \Bigl(1-H(\frac{1-d}{2})\Bigr) 
 \approx -\, d^2/(8\log 2) + d^2/(4\log 2), 
\end{equation*}
which is $d^2/(8\log 2)$.

\section*{Appendix B: Capacity Results for Binary Channels}

\begin{table}[h]
\caption{BA output results with non-symmetric binary channel, where $\epsilon=0.0001$
and the estimated formula (\ref{Formula_Est_nonsymmetric_binary_channel}) is used}
\label{T1}
\begin{center}
\renewcommand{\tabcolsep}{.5cm}
\renewcommand{\arraystretch}{1.5}   
\begin{tabular}{|c|c|c|c|} \hline 
$d$ & $C$ & $D_{1/2}(Q[0] \| Q[1])/2$ & Est.\\ \hline 
0.05 & 0.0003 & 0.0003 &  0.0005 \\ \hline 
0.10 & 0.0013 & 0.0013 &  0.0018 \\ \hline 
0.15 & 0.0028 & 0.0028 &  0.0041 \\ \hline 
0.20 & 0.0051 & 0.0051 &  0.0072 \\ \hline 
0.25 & 0.0080 & 0.0080 &  0.0113 \\ \hline 
0.30 & 0.0116 & 0.0116 &  0.0162 \\ \hline 
0.35 & 0.0159 & 0.0161 &  0.0221 \\ \hline 
0.40 & 0.0210 & 0.0213 &  0.0289 \\ \hline 
0.45 & 0.0270 & 0.0275 &  0.0365 \\ \hline 
0.50 & 0.0338 & 0.0347 &  0.0451 \\ \hline 
0.55 & 0.0417 & 0.0430 &  0.0546 \\ \hline 
0.60 & 0.0507 & 0.0527 &  0.0649 \\ \hline 
0.65 & 0.0610 & 0.0639 &  0.0762 \\ \hline 
0.70 & 0.0728 & 0.0771 &  0.0884 \\ \hline 
0.75 & 0.0865 & 0.0927 &  0.1014 \\ \hline 
0.80 & 0.1023 & 0.1116 &  0.1154 \\ \hline 
0.85 & 0.1211 & 0.1350 &  0.1303 \\ \hline 
0.90 & 0.1440 & 0.1657 &  0.1461 \\ \hline 
0.95 & 0.1737 & \textcolor{red}{0.2107} &  0.1628 \\ \hline 
\end{tabular}
\end{center}
\end{table}

\begin{table}[h]
\caption{BA output results with BSC, where $\epsilon=0.0001$,
the theoretical formula (\ref{Formula_theory_bsc})
and
the estimated formula $d^2/(2\log(2))$ are used
}
\label{T0}
\begin{center}
\renewcommand{\tabcolsep}{.5cm}
\renewcommand{\arraystretch}{1.5}   
\begin{tabular}{|c|c|c|c|c|} \hline 
$d$ & $C$ & Theory & $D_{1/2}(Q[0] \| Q[1])/2$ & Est.\\ \hline 
0.05 & 0.0012 & 0.0013 & 0.0013 & 0.0018 \\ \hline 
0.10 & 0.0050 & 0.0050 & 0.0050 & 0.0072 \\ \hline 
0.15 & 0.0113 & 0.0113 & 0.0114 & 0.0162 \\ \hline 
0.20 & 0.0201 & 0.0201 & 0.0204 & 0.0289 \\ \hline 
0.25 & 0.0316 & 0.0316 & 0.0323 & 0.0451 \\ \hline 
0.30 & 0.0457 & 0.0457 & 0.0472 & 0.0649 \\ \hline 
0.35 & 0.0626 & 0.0626 & 0.0653 & 0.0884 \\ \hline 
0.40 & 0.0823 & 0.0823 & 0.0872 & 0.1154 \\ \hline 
0.45 & 0.1050 & 0.1050 & 0.1131 & 0.1461 \\ \hline 
0.50 & 0.1308 & 0.1308 & 0.1438 & 0.1803 \\ \hline 
0.55 & 0.1600 & 0.1600 & 0.1801 & 0.2182 \\ \hline 
0.60 & 0.1927 & 0.1927 & 0.2231 & 0.2597 \\ \hline 
0.65 & 0.2294 & 0.2294 & 0.2745 & 0.3048 \\ \hline 
0.70 & 0.2704 & 0.2704 & 0.3367 & 0.3535 \\ \hline 
0.75 & 0.3164 & 0.3164 & 0.4133 & 0.4058 \\ \hline 
0.80 & 0.3681 & 0.3681 & 0.5108 & 0.4617 \\ \hline 
0.85 & 0.4268 & 0.4268 & 0.6410 & 0.5212 \\ \hline 
0.90 & 0.4946 & 0.4946 & \textcolor{red}{0.8304} & 0.5843 \\ \hline 
0.95 & 0.5762 & 0.5762 & \textcolor{red}{1.1640} & 0.6510 \\ \hline 
\end{tabular}
\end{center}
\end{table}

\end{document}